\documentclass[a4paper]{jpconfmodified}
\usepackage{epsfig,graphicx,url,twoopt}
\usepackage[numbers]{natbib}
\usepackage{hyperref}  
\usepackage{breakurl}

\newcount\longrefs
\def\aap{\ifnum\longrefs=1 {Astron.\ Astrophys.}\else 
                           {A\hbox{\rm \&}A}\fi}
\def\aapr{\ifnum\longrefs=1 {Astron.\ Astrophys.\ Rev.}\else 
                            {A\hbox{\rm \&}AR}\fi}
\def\aaps{\ifnum\longrefs=1 {Astron.\ Astrophys.\ Suppl.}\else 
                            {A\hbox{\rm \&}A Suppl.}\fi}
\def\aipcs{\ifnum\longrefs=1 {Am.\ Inst.\ Phys.\ Conf.\ Series}\else
                             {AIP Conf.\ Ser.}\fi}
\def\aj{\ifnum\longrefs=1 {Astron.\ J.}\else 
                          {AJ}\fi} 
\def\ao{\ifnum\longrefs=1 {Applied Optics}\else 
                           {Appl.\ Opt.}\fi} 
\def\aspcs{\ifnum\longrefs=1 {Astron.\ Soc.\ Pacific Conf.\ Series}\else 
                           {ASP Conf.\ Ser.}\fi} 
\def\apj{\ifnum\longrefs=1 {Astrophys.\ J.}\else 
                           {ApJ}\fi} 
\def\apjl{\ifnum\longrefs=1 {Astrophys.\ J. Lett.}\else 
                            {ApJ}\fi} 
\def\aplett{\ifnum\longrefs=1 {Astrophys.\ J. Lett.}\else 
                            {ApJ}\fi} 
\def\apjs{\ifnum\longrefs=1 {Astrophys.\ J. Suppl.}\else 
                            {ApJS}\fi}
\def\apss{\ifnum\longrefs=1 {Astrophys.\ and Space Science}\else 
                            {Astrophys.\ Space Sci.}\fi}
\def\araa{\ifnum\longrefs=1 {Ann.\ Rev.\ Astron.\ Astrophys.}\else 
                            {ARA\hbox{\rm \&}A}\fi}
\def\azh{\ifnum\longrefs=1 {Astronomicheskii Zhurnal}\else 
                            {Astron.\ Zhur.}\fi}
\def\baas{\ifnum\longrefs=1 {Bull.\ Am.\ Astron.\ Soc.}\else 
                            {BAAS}\fi}
\def\bain{\ifnum\longrefs=1 {Bull.\ Astronom.\ Institutes Netherlands}\else
                            {Bull.\ Astr.\ Inst.\ Neth.}\fi}
\def\gca{\ifnum\longrefs=1 {Geochim.\ Cosmochim.\ Acta}\else 
                           {Geochim.\ Cosmochim.\ Acta}\fi}
\def\grl{\ifnum\longrefs=1 {Geophys.\ Res.\ Lett.}\else 
                           {Geoph.\ Res.\ Lett.}\fi}
\def\iaucirc{\ifnum\longrefs=1 {IAU Circulars}\else 
                          {IAU Circ.}\fi}
\def\ip{\ifnum\longrefs=1 {in press}\else 
                          {in press}\fi}
\def\jgr{\ifnum\longrefs=1 {J.\ Geophys.\ Res.}\else 
                           {J.\ Geophys.\ Res.}\fi}  
\def\jrasc{\ifnum\longrefs=1 {J.\ Royal Astron.\ Soc.\ Canada}\else 
                           {JRAS Can.}\fi}  
\def\memsai{\ifnum\longrefs=1 {Mem.~Soc.~Astron.~Italiana}\else
                              {MemSAI}\fi}
\def\mnras{\ifnum\longrefs=1 {Mon.\ Not.\ Roy.\ Astron.\ Soc.}\else 
                             {MNRAS}\fi} 
\def\nat{\ifnum\longrefs=1 {Nature}\else 
                           {Nat}\fi}
\def\pasj{\ifnum\longrefs=1 {Pub.\ Astron.\ Soc.\ Japan}\else 
                            {PASJ}\fi} 
\def\pasp{\ifnum\longrefs=1 {Pub.\ Astron.\ Soc.\ Pacific}\else 
                            {PASP}\fi} 
\def\physscr{\ifnum\longrefs=1 {Physica Scripta}\else 
                            {Phys.\ Scrip.}\fi} 
\def\planss{\ifnum\longrefs=1 {Planetary \& Space Science}\else 
                            {Plan. \& Space Sci.}\fi} 
\def\procspie{\ifnum\longrefs=1 {Proc.\ SPIE}\else 
                            {Proc.\ SPIE}\fi} 
\def\qjras{\ifnum\longrefs=1 {Quarterly J.\ Royal Astron.\ Soc.}\else 
                            {QJRAS}\fi} 
\def\sa{\ifnum\longrefs=1 {Soviet Astron..}\else 
                               {Sov.\ Astron.}\fi}
\def\skytel{\ifnum\longrefs=1 {Sky \& Telescope}\else 
                            {Sky \& Tel.}\fi} 
\def\solphys{\ifnum\longrefs=1 {Solar Phys.}\else 
                               {Sol.\ Phys.}\fi}
\def\sovast{\ifnum\longrefs=1 {Soviet Astronomy}\else 
                               {Sov.\ Ast.}\fi}
\def\ssr{\ifnum\longrefs=1 {Space Science Rev.}\else 
                               {Space\ Sci.\ Rev.}\fi}
\def\zap{\ifnum\longrefs=1 {Zeitschr.\ f.\ Astrophysik}\else
                               {Z.\ Astrophys.}\fi}

\def\nl{,\ } %%\def\nl{\newline}  %% redefine as \newline for mail addresses

\def\IAC{Instituto de Astrof{\'{\i}}sica de Canarias\nl C/ Via Lactea S/N\nl
         E--38200 La Laguna, Tenerife\nl Spain}

\def\ISP{Institute for Solar Physics\nl 
         Albanova University Center\nl
         SE--106 91 Stockholm, Sweden}
\def\ITA{Institute of Theoretical Astrophysics\nl
         University of Oslo\nl
         P.O. Box 1029, Blindern\nl N--0315 Oslo\nl Norway}

\def\LA{Lingezicht Astrophysics\nl 't Oosteneind 9\nl 4158\,CA Deil\nl 
        The Netherlands}

\def\specchar#1{\uppercase{#1}}    %% to be redefined for A&A, small caps
\def\CIV{\mbox{C\,\specchar{iv}}} 
 
\def\CaII{\mbox{Ca\,\specchar{ii}}}

\def\FeIX{\mbox{Fe\,\specchar{ix}}}

\def\Hmin{\hbox{\rmH$^{^{_-}}\!$}}      %% H^min, very elegant
\def\HeI{\mbox{He\,\specchar{i}}}

\def\MgII{\mbox{Mg\,\specchar{ii}}}

\def\MnI{\mbox{Mn\,\specchar{i}}}

\def\Halpha{\mbox{H\hspace{0.1ex}$\alpha$}} %% \Halpha\ for space behind it

\def\Hbeta{\mbox{H\hspace{0.2ex}$\beta$}}
\def\Hgamma{\mbox{H\hspace{0.2ex}$\gamma$}}

\def\Lyalpha{\mbox{Ly$\hspace{0.2ex}\alpha$}}

\def\NaD{\mbox{Na\,\specchar{i}\,D}}    %% use \NaD\ for space behind it

\def\CaIIH{\mbox{Ca\,\specchar{ii}\,\,H}}

\def\HK{\mbox{H\,\&\,K}}
\def\hk{\mbox{h\,\&\,k}}

\def\level #1 #2#3#4{$#1 \; ^{#2} \mbox{#3} ^{#4}$}

 \def\rmH{{\rm H}}

\def\is{\!=\!}                             %% tighter spacing
\def\={\hbox{$\!=\!$}}                     %% no space around =

\def\spacing #1{\small\renewcommand{\baselinestretch}{#1}\normalsize}
\def\figspath{/home/rutten/rr/wrt/papers/rutten/2012-palmcove/figs/}

\def\figspath{}
\makeatletter  %RR \citeads for this numerical reference style
 \newcommandtwoopt{\citeads}[3][][]{\href{http://adsabs.harvard.edu/abs/#3}%
   {\def\hyper@linkstart##1##2{}%
    \let\hyper@linkend\@empty\cite[#1][#2]{#3}}}
\makeatother
\begin{document}

\title{Ellerman bombs: fallacies, fads, usage}

\author{Robert J. Rutten$^{1,2}$, 
        Gregal J. M. Vissers$^2$, 
        Luc H. M. Rouppe van der Voort$^2$, 
        Peter S\"utterlin$^3$ 
        and 
        Nikola Vitas$^4$}

\address{$^1$~\LA}
\address{$^2$~\ITA}
\address{$^3$~\ISP}
\address{$^4$~\IAC}

\ead{R.J.Rutten@uu.nl}

\begin{abstract}
  Ellerman bombs are short-lived brightenings of the outer wings of
  \Halpha\ that occur in active regions with much flux emergence.
  We point out fads and fallacies in the extensive Ellerman bomb
  literature\footnote{Clicking on citation numbers should open the
    corresponding ADS abstract page in a browser.}, 
%RR footnote only in astroph
  discuss their appearance in various spectral diagnostics, and
  advocate their use as indicators of field reconfiguration in
  active-region topography using AIA 1700\,\AA\ images.
\end{abstract}

%%%%%%%%%%%%%%%%%%%%%%%%%%%%%%%%%%%%%%%%%%%%%%%%%%%%%%%%%%%%%%%%%%%%%%%%%%%%
\section{Introduction}
%%%%%%%%%%%%%%%%%%%%%%%%%%%%%%%%%%%%%%%%%%%%%%%%%%%%%%%%%%%%%%%%%%%%%%%%%%%%
\spacing{0.99} %RR to get conclusion on one page
Ellerman \citeads{1917ApJ....46..298E} % Ellerman
described ``solar hydrogen bombs'' in 1917 as intense brightenings of
the extended wings of \Halpha, \Hbeta\ and \Hgamma, not visible in
other lines and with the line cores unaffected.
They last a few minutes and occur repetitively in active regions with
much flux emergence, preferentially near and especially between
penumbrae.
These properties define the Ellerman bomb (EB) phenomenon. The
subsequent EB literature cannot be fully reviewed here but we point
out some fads and fallacies below.
Bray and Loughhead
\citeads{1974soch.book.....B} % Bray+Loughhead Chromosphere
concluded in 1974 that {\em ``extensive modern observations have added
  little to Ellerman's original description''\/}.
This lack of progress changed with high-resolution observing, first in
the \href{http://sd-www.jhuapl.edu/FlareGenesis}{Flare Genesis}
flight \citeads{2002SoPh..209..119B} % Bernasconi++ Flare Genesis
\citeads{2002ApJ...575..506G} % Georgoulis++ Flare Genesis
\citeads{2002ESASP.506..911S} % Schmieder++
\citeads{2004ApJ...601..530S} % Schmieder++
\citeads{2004ApJ...614.1099P} % Pariat++
\citeads{2006AdSpR..38..902P} % Pariat++ Flare Genesis, Omega emergence
%##
and more recently with the
\href{http://www.solarphysics.kva.se}{Swedish 1-m Solar Telescope}
(SST) \citeads{2011ApJ...736...71W} % Watanabe++ EB1
\cite{Vissers++2013a}. % EB2
%####

%============================================================================
\begin{figure*}
  \centering
  \includegraphics[width=12cm]{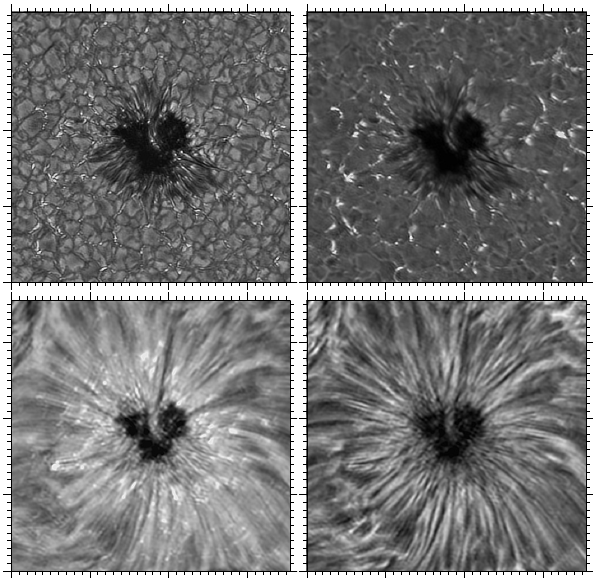}
  \caption[]{\label{fig:ebduds} %
    Simultaneous DOT images of the main spot of decaying AR\,10789 at
    view angle $\mu \is 0.86$.
    Clockwise: G band; blue \CaIIH\ wing at $\Delta \lambda \is
    -2.35$~\AA; sum of the \Halpha\ wings at $\Delta \lambda \is \pm
    0.5$~\AA; \Halpha\ core.
    Tick marks: arcseconds.}
\end{figure*}
%===========================================================================

%============================================================================
\begin{figure*}
  \centering
  \includegraphics[width=12cm]{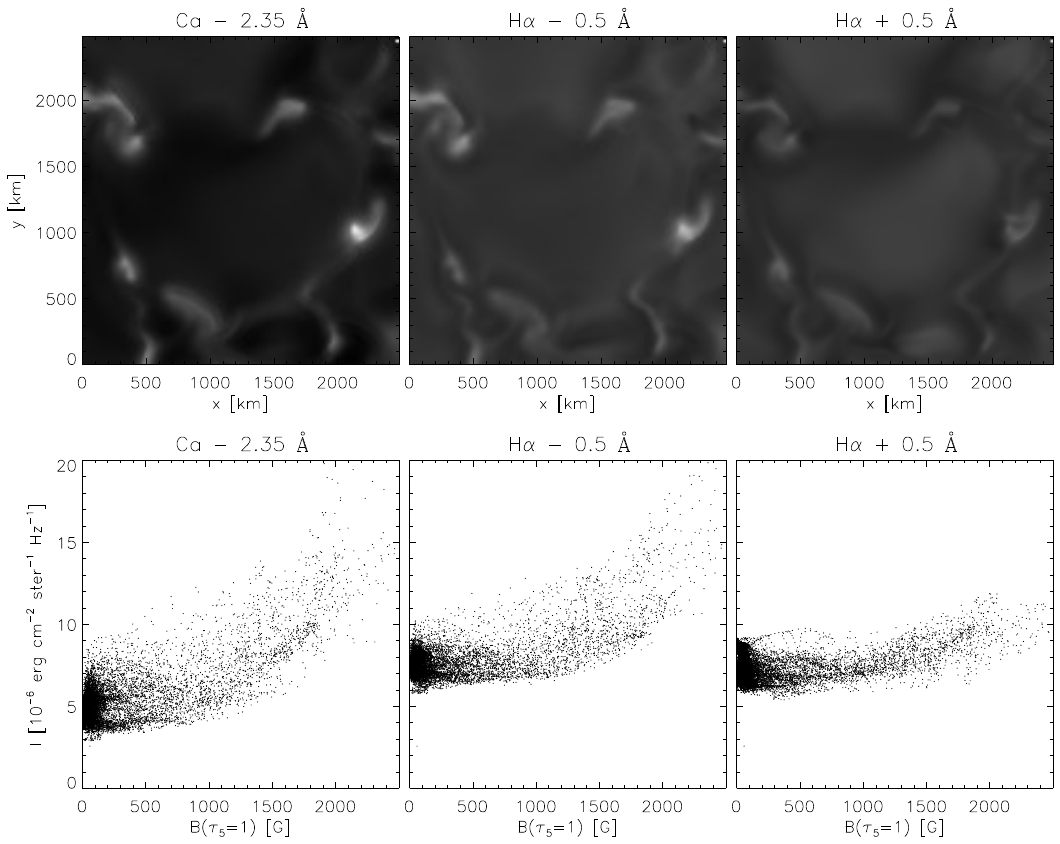}
  \caption[]{\label{fig:muram} %
    Synthesis of \CaIIH\ wing and \Halpha\ wing images for the
    numerical MHD simulation snapshot of
    \citeads{2009A&A...499..301V}. % Vitas++ MnI
    The scatter plots show brightness against magnetic field strength
    per pixel.
    Downdrafts make some MCs, including the brightest, brighter in the
    blue \Halpha\ wing.}
\end{figure*}
%===========================================================================

%%%%%%%%%%%%%%%%%%%%%%%%%%%%%%%%%%%%%%%%%%%%%%%%%%%%%%%%%%%%%%%%%%%%%%%%%%%%
\section{Magnetic concentrations as pseudo Ellerman bombs}
%%%%%%%%%%%%%%%%%%%%%%%%%%%%%%%%%%%%%%%%%%%%%%%%%%%%%%%%%%%%%%%%%%%%%%%%%%%%
Figure~\ref{fig:ebduds} shows snapshots from 3-hour multi-wavelength
\href{http://www.staff.science.uu.nl/~rutte101/dot}{Dutch Open
  Telescope} (DOT)
movies\footnote{Links: \href{http://www.staff.science.uu.nl/~rutte101/dot/albums/movies/2005-07-13-AR10789-gb+caw+haw+hac.mov}{four-panel movie},
  % {\url{http://www.staff.science.uu.nl/~rutte101/dot/albums/movies/2005-07-13-AR10789-gb+caw+haw+hac.mov}}
  \href{http://www.staff.science.uu.nl/~rutte101/dot/albums/movies/2005-07-13-AR10789-habw.mov}{wing-only movie}.}.
  % {\url{http://www.staff.science.uu.nl/~rutte101/dot/albums/movies/2005-07-13-AR10789-habw.mov}}.}.
The \Halpha\ wing image in the lower-left panel shows many bright
grains that in the movie adhere to Ellerman's description: {\em ``they
  seem to follow one another like the balls of a Roman candle''\/}.
About half exceed the mean intensity by over 30\%, a few are brighter
than 54\% (4\,$\sigma$) excess.
Traditional EB threshold criteria
%% \citeads{2002ApJ...575..506G} % Georgoulis++ Flare Genesis
might classify these as EBs.
However, movie inspection including comparison with the parallel
\CaIIH\ and G-band movies shows that all the bright grains are simply
``network bright points'' marking strong-field magnetic concentrations
(MCs).
They are not EBs but ``moving magnetic features'' described by
\citeads{1969SoPh....9..347S} % Sheeley
\citeads{1971IAUS...43..329V} % Vrabec also MCs
\citeads{1973SoPh...28...61H} % Harvey+Harvey
\citeads{1974IAUS...56..201V} % Vrabec
\citeads{1987SoPh..112..295M} % Muller+Mena decaying spot
\citeads{2005ApJ...635..659H}. % Hagenaar+Shine
Nor was there flux emergence in this region evident in
\href{http://sohowww.nascom.nasa.gov}
{SOHO}/\href{http://soi.stanford.edu}{MDI} magnetograms.
%RR waarom werken deze weblinks niet in acroread voor de pdf?

MCs are bright in the continuum from hot-wall radiation
\citeads{1976SoPh...50..269S}. % Spruit hot wall heating
The classic magnetostatic thin fluxtube/sheet paradigm
\citeads{1977PhDT.......237S} % Spruit thesis
\citeads{1981NASSP.450..385S} % Spruit fluxtubes Sun-as-Star
\citeads{1984A&A...139..426D} % Deinzer++ KIS flux sheets
\citeads{1988A&A...194..257K} % Knoelker++ KIS flux sheets
\citeads{1991sia..book..890S} % Spruit++ fluxtubes Sol-int+atm
\citeads{1992A&A...262L..29S} % Solanki+Brigljevic fluxtubes
\citeads{1993A&A...268..736B} % Bunte+Solanki+Steiner wine glass model
\citeads{1993SSRv...63....1S} % Solanki new testament
explains their physics well and is well applicable
\citeads{2009A&A...504..595Y}, % Yelles++ thin fluxtube vs 3D
although actual MCs show complex morphology with rapid changes
\citeads{2004A&A...428..613B} % Berger++, BP's = flowers
\citeads{2005A&A...435..327R}. % Rouppe+ flower ribbon evolution
Since MCs are small and reside in dark intergranular lanes, their
continuum brightening is only seen at sub-arcsecond resolution
\citeads{1996ApJ...463..797T}. % Title+Berger G BP double Gauss
MCs also appear as bright ``line gaps'' in the cores of neutral-metal
lines due to ionization from evacuation, particularly in \MnI\ lines
\citeads{1987ApJ...314..808L} %C Livingston++ luminosity variation 
\citeads{2007ApJ...657.1137L} %C Livingston++ flux cycle variations
for which the contrast is not weakened by Dopplershifts in the
surrounding granulation \citeads{2009A&A...499..301V}. % Vitas++ MnI
They also appear markedly bright in the outer wings of \Halpha\
\citeads{2006A&A...449.1209L}, % Leenaarts Halpha BPs
at larger contrast than in other diagnostics including the G band
\citeads{2006A&A...452L..15L}. % Leenaarts++ BP diagnostics
For the G band the contrast enhancement comes from dissociation of the
% ##
CH molecules producing this feature, for the \Halpha\ wings from
smaller collisional damping.
Both result also from MC evacuation.

The \Halpha\ core image in Fig.~\ref{fig:ebduds} shows the
chromospheric fibril canopy that overlies and shields the deep
photosphere imaged in the far \Halpha\ wings.
\Halpha\ does not sample the intermediate layers (seen in \CaIIH\ as
shock-ridden clapotisphere
\citeads{2012RSPTA.370.3129R}) % Rutten RoySoc review
due to its low-temperature opacity gap
\citeads{1972SoPh...22..344S} % Schoolman
\citeads{2012A&A...540A..86R} % Rutten+Uitenbroek popdeps Mg+Ha
\citeads{2012ApJ...749..136L}, % Leenaarts++ Halpha
so that at any wavelength its formation jumps between chromosphere and
deep photosphere depending on the chromospheric fibril opacity.
The $\Delta \lambda \is \pm 0.5$~\AA\ sampling wavelength of
Fig.~\ref{fig:ebduds} mixes deep-photosphere and upper-chromosphere
brightness contributions, in this case MCs and superpenumbral fibrils.
The fibrils look like the line-core ones but are thinner.
The MCs show about the same morphology as in the \CaII-wing and G-band
images because their brightness originates at about the same depth.
The overlying fibrils block some MCs.
Where they are sufficiently thin that the photospheric MC brightness
shines through, they degrade the MC image sharpness.
This defocus, compared to the sharp LTE-formed G-band and \CaIIH-wing
brightness features, results from ray spreading across the opacity gap
and scattering through the effectively thin fibrils.

Note in the \Halpha\ wing image that the shielding by overlying
fibrils appears to be less for the MCs in the moat than for MCs
further away.
%##
Larger transparency of superpenumbral fibrils may result from
repetitive shock heating by running penumbral waves
\citeads{1972ApJ...178L..85Z} % Zirin+A.Stein discovery RPW
\citeads{2001A&A...375..617C}, % C+G+K RPWs
ionizing hydrogen too frequently to let it reach population equilibrium
since hydrogen ionization/recombination balancing is exceptionally
slow in cooling shock aftermaths due to the large $n$\,=\,1--2
\Lyalpha\ transition
\citeads{2002ApJ...572..626C} % Carlsson+Stein dynamic H ionization
\citeads{2007A&A...473..625L}. % Leenaarts++ non-E hydrogen
%####

In the
\href{http://www.staff.science.uu.nl/~rutte101/dot/albums/movies/2005-07-13-AR10789-habw.mov}{DOT
  movie} some of the larger MCs do brighten momentarily, especially in
the blue \Halpha\ wing.
Blue-wing brightening may result from MC downdrafts, as in the
numerical simulation shown in Fig.~\ref{fig:muram}.
Downdrafts often occur in moat MCs that have opposite polarity to the
sunspot \citeads{2007A&A...471.1035Z}. % Zhang++ MDI
MC downdrafts tend to produce shocks higher up
\citeads{2011ApJ...730L..24K} % Kato+Steiner++ downdrafts-slow-modes-shocks
that are best seen in \NaD\ Dopplergrams
\citeads{2011A&A...531A..17R}. % Rutten++ NaD1 revgran
Inspection of simultaneous \Halpha\ and \NaD\ active-region
Dopplergram sequences from the SST shows that such MC shock occurrence
is often accompanied by \Halpha\ blue-wing brightening.
%RR @ demonstrate in superpen paper

MC brightening may also result from field concentration by bathtub
vorticity in granular swirls
\citeads{2008ApJ...687L.131B}. % Bonet++ vortex
Such brightening also reflects increasing hot-wall radiation, and must
not be misinterpreted as MHD heating or reconnection
\citeads{2007A&A...476..971J}. % Jess++ vortex = reconnection
However, swirl brightening seems a rare phenomenon.
In the DOT data of Fig.~\ref{fig:ebduds} \Halpha\ wing brightening
shows no correlation with vorticity in the granular flows measured
from the G-band movie.
Rather, it happens in concert with \CaIIH-wing brightenings that seem
mostly due to episodes of magnetic patch
\citeads{2005A&A...441.1183D} % de Wijn++ patches
compression in the outward moat flow.

Upshot regarding EB fallacies: part of the EB literature (and the
moustache literature, Severny's
\citeads{1956Obs....76..241S} % Severny fine structure = moustaches.
term for far-extended bright line wings) concerns MCs, not EBs.
This is the case when dense network was not resolved, moustache
brightening of the continuum and neutral-metal lines was observed, or
when low \Halpha-wing contrast thresholds were applied.
In our opinion this warning should be heeded when reading
\citeads{1956Obs....76..241S} % Severny fine structure = moustaches.
\citeads{1957SvA.....1..668S} % Severny review moustaches
\citeads{1968IAUS...35..293B} % Bruzek bright points
\citeads{1970SoPh...14..112T} % Thuron+Lena moustaches 10 mu
\citeads{1972SoPh...26...94B} % Bruzek moustaches %ebduds 80%
\citeads{1987SoPh..108..227Z} % Zacharadias++ EBs in Halpha
\citeads{1991A&A...251..675S} % Stellmacher+Wiehr on BP moustaches
\citeads{1995A&A...296..567D} % Denker++ moustaches speckle (BPs?)
\citeads{1997A&A...323..599D} % Denker moustaches (= BPs)
\citeads{1998SoPh..182..381N} % Nindos+Zirin EBs
\citeads{1999ESASP.448..279G} % Georgakilas+Christopulou+Koutchmy
\citeads{2013SoPh..283..307N}. % Nelson++ EBs

%============================================================================
\begin{figure*}
  \centering
  \includegraphics[width=14cm]{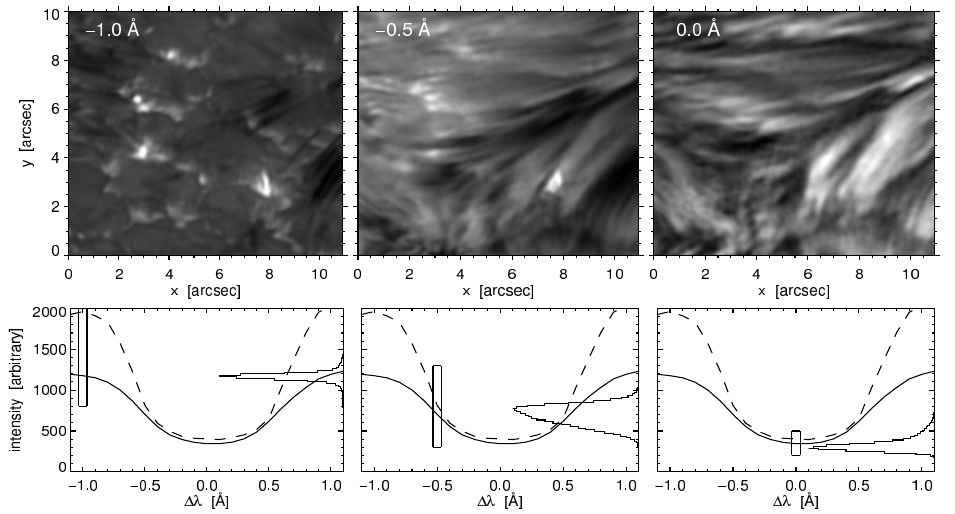}
  \caption[]{\label{fig:EB1} %
    Upper row: SST EB images in \Halpha\ at the specified wavelength
    separations from line center, taken from Fig.~1 of
    \citeads{2011ApJ...736...71W}. % Watanabe++ EBs
    View angle $\mu \is 0.67$ with the limb direction to the top.
    Lower row: \Halpha\ profile (solid) averaged over the full $67
    \times 67$~arcsec$^2$ field of view of the observations and
    \Halpha\ profile (dashed) for a pixel in the EB at $(x,y)\is
    (8.0,2.5)$.
    The histograms show the normalized intensity occurrence
    distribution over the full observed field of view at each
    wavelength.
    The rectangle widths show the passband FWHM, the vertical extents
    the greyscale clipping of the corresponding images (each
    bytescaled for maximum contrast).
  }
\end{figure*}
%===========================================================================

%%%%%%%%%%%%%%%%%%%%%%%%%%%%%%%%%%%%%%%%%%%%%%%%%%%%%%%%%%%%%%%%%%%%%%%%%%%%
\section{Ellerman bombs are not chromospheric}
%%%%%%%%%%%%%%%%%%%%%%%%%%%%%%%%%%%%%%%%%%%%%%%%%%%%%%%%%%%%%%%%%%%%%%%%%%%%
Figure~\ref{fig:EB1} after Fig.~1 of
\citeads{2011ApJ...736...71W} % Watanabe++ EBs
shows EBs at the unprecedented resolution of the SST.  In this slanted
limbward high-resolution view EBs appear in the outer \Halpha\ wings
as tiny bright upright flames that flicker rapidly while their
footpoints travel along MC-filled network strands.  The ensemble may
be bright for many minutes but the subflames last only seconds.
Inspection of the corresponding SST
movies\footnote{\href{http://iopscience.iop.org/0004-637X/736/1/71/fulltext}
  {\url{http://iopscience.iop.org/0004-637X/736/1/71/fulltext}};
  \href{http://www.staff.science.uu.nl/~rutte101/rrweb/rjr-talkstuff/bright-points-flaming.mpg}
  {one example}.}  demonstrates this behavior irrevocably.  The
elongated shape was noted before in lower-resolution data
\citeads{1982SoPh...79...77K} % Kurokawa++
\citeads{2002ApJ...575..506G} % Georgoulis++ Flare Genesis
\citeads{2007A&A...473..279P}, % Pariat++
as was their intermittent substructure
\citeads{2010PASJ...62..879H}. % Hashimoto++ substructure EBs

Figure~\ref{fig:EB1} also demonstrates that EBs are purely
photospheric phenomena.
They are not seen at \Halpha\ line center because the flames, even
when tall (a few hundred to a thousand km
\citeads{1963Obs....83...37H} % Harvey EB height off limb 400 km
\citeads{1982SoPh...79...77K} % Kurokawa++
\citeads{2008ApJ...684..736W}), % Watanabe++ EBs
do not break through the overlying dense canopy of chromospheric
\Halpha\ fibrils that always covers a growing active region.
Comparison of the greyscale ranges (narrow rectangles in the lower
panels of Fig.~\ref{fig:EB1}) shows that the EB wing emission is much
brighter than the brightest line-core features.
Some diffuse line-center brightening might again result from
photospheric EB emission that passes through the opacity gap and
scatters through the fibrils, but our SST movies suggest that such
line-center brightening above EBs is uncommon.
At $\Delta \lambda \is -0.5$~\AA\ the fibrils show the same morphology
as at line center but they are less opaque and therefore brighter
\citeads{2012ApJ...749..136L}, % Leenaarts++ Halpha
except for blueshifted dark ones seen also at $\Delta \lambda \is
-1.0$~\AA.
At $\Delta \lambda \is -0.5$~\AA\ the EBs do shine through, again with
loss of sharpness from scattering.
Some MCs also shine through, but less brightly than in
Fig.~\ref{fig:ebduds} because the bytescale now includes much brighter
EBs and the \Halpha\ fibril canopy is thicker in this flux-emergence
region than around the little decaying spot of Fig.~\ref{fig:ebduds}.

%##
Thus, an EB top may reach higher than the 400--500~km nominal height
of the temperature minimum in standard one-dimensional
static-equilibrium models of the solar atmosphere
\citeads{1971SoPh...18..347G} % HSRA
\citeads{1981ApJS...45..635V} % VALIII
\citeads{1993ApJ...406..319F} % FAL including FALC 
\citeads{2008ApJS..175..229A}. % Avrett+Loeser update FALC SUMER
The onset of the actual solar outward temperature rise likely varies
between 200~km in MCs and 2000~km in internetwork, varying temporally
as well \citeads{2007A&A...473..625L} % Leenaarts++ non-E hydrogen
\citeads{2010ApJ...709.1362L}, % Leenaarts++ NaD1
but in any case, the jet-like EB flame protrusions originate from the
deep photosphere, in the network at the EB footpoint, and do not
affect or poke through the overlying chromosphere defined
\citeads{2012RSPTA.370.3129R} % Rutten RoySoc review
by the \Halpha\ fibrils.
%####

Apparent blue-brighter-than-red EB wing asymmetry
\citeads{1970SoPh...11..276K} % Koval+Severny moustache asymmetry
usually results from inverse Evershed flows
\citeads{1962AuJPh..15..327B} % Beckers invEv flow
\citeads{1990A&A...233..207D} % Dere++ revEv flow UV and Halpha
\citeads{2008A&A...491L...5T} % Teriaca++ invEv in TR
along the fibrils, the dark chromospheric line core shifting into the
red-wing emission.
Thus, asymmetries and wavelengths of EB emission peaks do not define
Dopplershifts of the EB emitting material
\citeads{1999ESASP.448..279G} % Georgakilas+Christopulou+Koutchmy
\citeads{2004ApJ...601..530S} % Schmieder++
but are set by the absorbing overlying fibrils
%%\citeads{1972SoPh...26...94B} % Bruzek moustaches # mostly MCs 
\citeads{1997A&A...322..653D}. % Dara++

\begin{figure*}
  \centering
  \mbox{~~~~~~}\includegraphics[width=15cm]{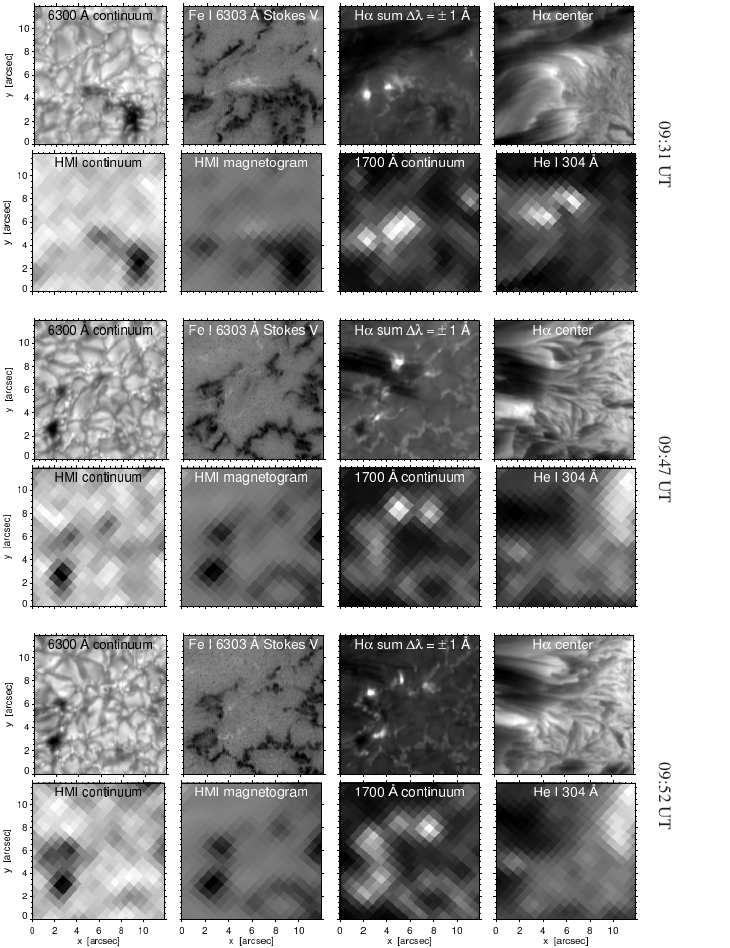} 
  \caption[]{\label{fig:EB2} %
    Two small-field cutouts of corresponding SST images (upper row of
    each set) and SDO images (lower rows) from the data of
    \cite{Vissers++2013a}. % in cdrjrbib temp.bib
    Magnification per pdf viewer is recommended for the SST images.
    View angle $\mu\is 0.89$; the panel tops are in the limb direction.   
    The third set shows the same cutout as the second, but five minutes later.
    Each panel is bytescaled independently. 
%% Ha-wing differs between the two sampling times of fov3 due to
%% brighter EBs
%##
%%    The HMI magnetograms are reversed in sign.  %RR SST convention opposite?
%####
    The arrows overlaid on the SST magnetograms show surface flows
    measured from the continuum sequence by correlation tracking.
    
  }
\end{figure*}

It has been suggested that EBs are accompanied by \Halpha\ surges
\citeads{1973SoPh...28...95R} % Roy surges emanate from EBs
\citeads{1973SoPh...30..449R} % Roy+Leparskas EB statistics
\citeads{1979SoPh...63..353C} % Carlqvist surges, moustaches flarelike
\citeads{1982SoPh...77..121S} % Shibata++ simulation of reconnection
\citeads{2007ApJ...657L..53I} % Isobe+Tripathi+Archontis
\citeads{2008PASJ...60...95M}. % Matsumoto++
However, the high-resolution SST data of
\citeads{2011ApJ...736...71W} % Watanabe++ EBs
contained only two tentative cases.
Our newer SST data sampled in Fig.~\ref{fig:EB2} contain at most one
questionable case \cite{Vissers++2013a}. % in cdrjrbib temp.bib
An EB--surge connection is certainly not ubiquitous
\citeads{2002ApJ...575..506G}. % Georgoulis++ Flare Genesis 
 
Upshot regarding EB fallacies: most EB papers err in describing the EB
phenomenon as chromospheric.  It is not.  
%RR "most papers" means no REF LIST here
EBs have no systematic counterpart in the overlying chromosphere,
transition region or corona.  Dark chromospheric \Halpha\ cores are
not EB ingredients, nor are their Dopplershifts.  

Comment regarding EB fads: EBs may combine strong downflows in the low
photosphere
\citeads{2007ASPC..369..113S} % Shimizu++ downflows from reconnection
with outward flows higher up
\citeads{2008PASJ...60...95M} % Matsumoto++
\citeads{2011ApJ...736...71W}. % Watanabe++ EB1
Such bi-directional flows are reminiscent of ``chromospheric anemone
jets'' seen in \href{http://solar-b.nao.ac.jp/index_e.shtml}{Hinode}
\CaIIH\ data
\citeads{2007Sci...318.1591S} % Shibata++ chrom anemone jets
\citeads{2010PASJ...62..901M} % Morita++ astroph
\citeads{2011ApJ...731...43N}. % Nishizuka++ anemone jets
If such \CaIIH\ jets actually are EBs, they are also confined to the
photosphere.  Just as for \Halpha, not every bright \CaIIH\ feature is
necessarily chromospheric.

%============================================================================
%===========================================================================

%%%%%%%%%%%%%%%%%%%%%%%%%%%%%%%%%%%%%%%%%%%%%%%%%%%%%%%%%%%%%%%%%%%%%%%%%%%%
\section{Ellerman bomb visibility}
%%%%%%%%%%%%%%%%%%%%%%%%%%%%%%%%%%%%%%%%%%%%%%%%%%%%%%%%%%%%%%%%%%%%%%%%%%%%
Even though they are photospheric, proper EBs (those that adhere to
Ellerman's description) do not show up in neutral-metal lines nor in
the continuum.  At sufficient angular resolution they do show
continuum moustaches, G-band brightening
\citeads{2011CEAB...35..181H}, % Herlender+Berlicki EB light curves
and narrow neutral-line gaps at their footpoint, sampling the MC from
which they arise.  Ellerman wrote: {\em ``they frequently appear in
  the faculae so that their spectra are superposed on those of
  faculae, thus giving the appearance of great extension to the bright
  ``bomb'' band\/''} which wasn't heeded in most pseudo-EB
literature.

%RR checked in 6300 cont, see cddata SST EB-2 tinker.pro
%RR clear in Rezaei talk FeI6303 spectrum

EB flames are also observable in \CaII\,8542\,\AA, \CaII\ \HK, and in
mid-UV continua
\citeads{2000ApJ...544L.157Q} % Qiu++ (BBSO) 2000 TRACE 1600 + Halpha EBs
\citeads{2002ApJ...575..506G} % Georgoulis++ Flare Genesis 
\citeads{2004ApJ...601..530S} % Schmieder++
\citeads{2006ApJ...643.1325F} % Fang++ Themis data + electron-beam models
\citeads{2006SoPh..235...75S} % Socas++ SPINOR CaII8542, downdrafts
\citeads{2007A&A...473..279P} % Pariat++ 8542, TRACE UV
\citeads{2008PASJ...60..577M} % Matsumoto++ Hinode CaIIH + Hida Halpha
\citeads{2010MmSAI..81..646B} % Berlicki+Heinzel+Avrett DOT+1600
\citeads{2011CEAB...35..181H} % Herlender+Berlicki
\cite{Vissers++2013a}. % in rjrbib temp.bib
They are not identically the same in these different diagnostics, but
they often show up at the same space-time locations.
We illustrate this for the 1700\,\AA\ continuum in Fig.~\ref{fig:EB2}
by combining image cutouts from the new SST data of
\cite{Vissers++2013a} % in rjrbib temp.bib
with co-aligned image cutouts from SDO.
Clearly, AIA's 1700\,\AA\ shows the larger EBs also, at much lower
resolution but with less interference by chromospheric fibrils.
%##
Since this continuum originates in the upper photosphere
\citeads{1981ApJS...45..635V} % VALIII
\citeads{2005ApJ...625..556F}, % Fossum+Carlsson 1700-1600 phase
the bottom part of an \Halpha\ EB is hidden in slanted 1700\,\AA\
viewing but this is not noticeable at AIA's angular resolution.
%####

AIA's 1600\,\AA\ images (not shown) show the EBs too, at yet larger
contrast, but these also show a few more extended and yet brighter
transition-region transients for this area and period.
They have short-loop morphology and correspond to bright patches in
\Halpha\ line-center, are also seen in \HeI\,304\,\AA, and sometimes
in \FeIX-X 171\,\AA.
These loop-like bright transients appear unrelated to underlying EBs.
We attribute their visibility in the 1600\,\AA\ images to the \CIV\
lines in this passband and tentatively identify them as
upper-atmosphere transients such as small flaring arch filaments and
microflares visible in transition-region diagnostics and \Halpha\ line
center.
The \HeI\,304\,\AA\ images in Fig.~\ref{fig:EB2} indeed mimic the
\Halpha\ line-center morphology which appears indifferent to EBs
underneath.

It is illustrative and recommended to play and blink SDO 1700\,\AA,
1600\,\AA, and \HeI\,304\,\AA\ movies of a very active region with
much flux emergence, for example AR\,11654 on January 10, 2013 and the
following days when it was crackling with short-lived brightenings in
these diagnostics.
The plethora of transient brightness features in 1600\,\AA\ are the
sum of different sorts, respectively seen in 1700\,\AA\ and in
\HeI\,304\,\AA.
Very short-lived, pointlike 1700\,\AA\ brightenings were likely
photospheric \Halpha-wing EBs, whereas more spread-out loop-like
longer-duration brightenings were likely flaring arch filaments and
small flares, bright in the transition-region diagnostics and probably
bright at \Halpha\ line center.
Inspection of such flux-emergence AIA movies and of our
high-resolution SST data suggests that distinction should be made
between point-like (at SDO resolution) EBs and loop-like flaring arch
filaments as different entities, respectively photospheric and
upper-atmosphere.
AIA's 1700\,\AA\ shows fewer of the latter type and is therefore the
best SDO diagnostic for EB studies.
Figure~\ref{fig:EB2} demonstrates that SDO's resolution is poor
compared with the SST's, but the obvious AIA advantage is that it
samples the stronger EBs in any Earth-side active region at any time.

Comment regarding EB fallacies: the low resolution of older data has
led to claims of apparent cospatiality of higher-atmosphere
brightenings with EBs.
Examples: coincidence with EBs was claimed for 6 out of 16 microflares
by
%%\citeads{1996SoPh..165..155K}, % Kitai+Muller WL-BPs and EBs
%RR OOPS wrong in final submission 
\citeads{2002ApJ...574.1074S} % Shimizu++ microflares
but the example pair in their Figs.~4 and 5 has 8~arcsecond
separation.
The \Halpha\ line-center brightening observed and modeled by
\citeads{2010ApJ...724.1083G} % Guglielmino++ EB like surge + simulation
was not an EB.
The $\Halpha\!\pm\!0.35$\,\AA\ feature claimed by
\citeads{2009ApJ...701..253M} % Madjarska++ explosive events
to be an EB at 1\,MK doesn't look like an EB to us.
The long-lived EB of
\citeads{2010CEAB...34...65H} % Herlender+Berlicki
at the foot of an arch filament looks like a large MC.
Similarly for the \CaIIH\ bright points bordering the flaring arch
filament in \citeads{2010ApJ...712L.111J}, % Jess++
much like the bright points of
\citeads{1996SoPh..165..155K}. % Kitai+Muller WL-BPs and EBs
\citeads{2004ApJ...601..530S} % Schmieder++
reported one coincidence of X-ray brightening at an exceptionally
bright EB but doubted ubiquitous co-spatiality.
We have aligned our new SST data with
\href{http://sdo.gsfc.nasa.gov}{SDO}/\href{http://aia.lmsal.com}{AIA}
image sequences and searched for but did not find any significant EB
impact on the overlying transition region (\HeI\,304\,\AA) and corona
(\FeIX\,171\,\AA) \cite{Vissers++2013a}. % in cdrjrbib temp.bib

Comment regarding EB fads: most EB studies are based on observations in
the \Halpha\ wings and typically describe a single, a few, or at most
some dozen EBs in a single active region during a short period.
This type of study should exploit \href{http://iris.lmsal.com}{IRIS}
in the near future since EBs will be well visible in \MgII\ \hk, as in
\CaII\ \HK.
On the other hand, it seems time to exploit the AIA~1700\,\AA\
database to study very many more with respect to occurrence patterns.

%%%%%%%%%%%%%%%%%%%%%%%%%%%%%%%%%%%%%%%%%%%%%%%%%%%%%%%%%%%%%%%%%%%%%%%%%%%%
\section{Ellerman bomb detection}
%%%%%%%%%%%%%%%%%%%%%%%%%%%%%%%%%%%%%%%%%%%%%%%%%%%%%%%%%%%%%%%%%%%%%%%%%%%%
In \citeads{2011ApJ...736...71W} % Watanabe++ EB1
EBs were identified and selected manually on the basis of their flame
morphology in the SST \Halpha-wing movies, meaning bright, narrow,
tall, upright, flickering appearance in the limbward view at this
unprecedented resolution.
More formal detection criteria are formulated for two SST data sets in
\cite{Vissers++2013a}. % EB2
One of the requirements defining an EB kernel is 55\% excess intensity
in the \Halpha\ wings over the spatial average of the active region,
while setting a threshold of 5\,$\sigma$ or higher above the mean
seems a good counterpart to select EBs in AIA 1700\,\AA\ images.
Both thresholds are passed by the EBs in Fig.~\ref{fig:EB2}, whereas
almost none of the bright \Halpha-wing features in the movie sampled
in Fig.~\ref{fig:ebduds} passed a 55\% \Halpha-wing excess threshold, nor a 
5\,$\sigma$ excess threshold in simultaneous 1600\,\AA\ images from
TRACE\footnote{Recently
  \citeads{2013SoPh..283..307N} % Nelson++ EBs
  claimed to detect over 3000 EBs during 90~minutes around a small spot
  rather like the one in Fig.~\ref{fig:ebduds}.
  Our application of the above threshold to the simultaneous SDO
  1700\,\AA\ images suggests that they mostly detected pseudo-EBs of
  which the \Halpha-wing brightness has nothing to do with MHD
  heating.}.
%RR /media/usb0/alldata/SDO/2010-11-18-nelson/record.idl

In addition, small spatial extent is a discriminator, as already
suggested by Ellerman: {\em ``On rarer occasions they [EBs] are
  superposed on bright reversals of \Halpha\ over eruptive regions,
  but this is an uncommon occurrence, and the distinction is easily
  made between the two phenomena by the flickering of the ``bomb''
  band compared to the \Halpha\ reversal, due to the effect of
  seeing.\/''} -- as using scintillation to distinguish stars from
planets.
Small size, fast variation, large brightness, and appearing in a
region with much flux emergence together become the recipe to locate
EBs in AIA 1700\,\AA\ image sequences.
We are presently refining such automated detection for multiple
viewing angles \cite{Vissers++2013b}. % EB3

%%%%%%%%%%%%%%%%%%%%%%%%%%%%%%%%%%%%%%%%%%%%%%%%%%%%%%%%%%%%%%%%%%%%%%%%%%%%
\section{Ellerman bomb radiation mechanism}
%%%%%%%%%%%%%%%%%%%%%%%%%%%%%%%%%%%%%%%%%%%%%%%%%%%%%%%%%%%%%%%%%%%%%%%%%%%%
In traditional one-dimensional stratification modeling along a
vertical line of sight the emergent \Halpha\ profile maps the
variation of the source function with depth, with the outermost wings
sampling the deepest layers
\citeads{2003rtsa.book.....R}. % Rutten RTSA notes
Postulating a suited temperature perturbation with appropriate
structure and motion to a standard model atmosphere can then explain
any \Halpha\ excess emission profile
\citeads{1983SoPh...87..135K} % Kitai 1D NLTE modeling
\citeads{2006ApJ...643.1325F}, % Fang++ Themis data + electron-beam models
but not unambiguously; different models may produce similar
\Halpha\ profiles
\citeads{2010MmSAI..81..646B}. % Berlicki+Heinzel+Avrett DOT pubs
%RR not bad but I doubt they do inelastic Thomson
 
In the slanted perspective of Figs.~\ref{fig:EB1} and \ref{fig:EB2}
the EB flames appear as extended, dense, hot slabs that stick up from
the network MCs into the otherwise \Halpha-transparent upper
photosphere,
%##
making cloud modeling more appropriate than one-dimensional
plane-parallel modeling.
%####
The EB excess emission and its profile including its
moustache extent are properties of such a slab, while the core
absorption and Dopplershift patterns are properties of the overlying
fibrils.  Further away from line center the slab is optically thinner
but not ``deeper''.

We attribute the non-visibilty of the EB slabs in
neutral-metal lines to neutral-metal ionization in the hot EB flames,
and similarly their transparency in the continuum to \Hmin\
ionization.
% RR in fluxtubes this isn't the case.  FeI ionizes in them too but
% they do not get very bright in the continuum.  But these are so deep
% that H is still 100 percent neutral anyhow, so that Hmin is only set
% by N_e.  With FeI etc ionization N_e gets larger so Hmin may
% actually increase a bit.  In EBs there should be appreciable H
% ionization so that N_e is set by H and no longer by metal
% ionization.
Considerable ionization of neutral hydrogen is also likely, making
cascade recombination the main producer of the EB \Halpha\ photons.
Even if the ionization occurs only briefly during a momentary reconnection
event, \Halpha\ will so shine in a longer-duration afterglow because
hydrogen recombination balancing is slow in the aftermath
\citeads{2002ApJ...572..626C} % Carlsson+Stein dynamic H ionization
\citeads{2007A&A...473..625L}. % Leenaarts++ non-E hydrogen

Following the suggestion of
\citeads{1968mmsf.conf..109E} % Engvold+Maltby
we attribute the extended bright moustaches to subsequent thermal
Thomson scattering of these photons, notwithstanding the small process
crosssection, because the flames combine high temperature with very
high (mid-photosphere) density.
While the line-center peak of the EB brightness profile remains hidden
by overlying fibrils, the wings may gain brightness at large thermal
electron Dopplershift whenever a line-core photon meets an electron
and is scattered our way.
%##
Since EBs are very optically thick at \Halpha\ line center, being
non-transparent already in the outer wings, we suggest that \Halpha\
resonance scattering confines line-core photons in a random walk
within the EB until they are Dopplershifted into a wing and escape.
This scattering mechanism will produce similar moustaches for other
chromospheric lines (principally \MgII\ \hk, \CaII\ \HK, and
\CaII\,8542\,\AA) if an EB similarly confines scattering line photons,
which is likely unless these ions ionize too much.
%####
Obviously, numerical simulation with detailed spectral synthesis
including non-monochromatic Thomson scattering may vindicate this
moustache mechanism.

In this view large moustache extent happens {\em because\/} EBs are
photospheric and makes them visible beyond the overlying fibrils in
the spectrum.
For \Halpha\ such scattering occurs only within the flame, there being
neither free electrons nor hydrogen atoms in the $n\is2$ level in the
surrounding atmosphere, and so it remains local.
%#### 
Flame images taken in the outer \Halpha\ wings therefore remain sharp,
as demonstrated in Figs.~\ref{fig:EB1} and \ref{fig:EB2}.
In \CaII\ \HK\ EBs get a diffuse halo from additional resonance
scattering in the surrounding upper photosphere
\citeads{2008PASJ...60..577M}. % Matsumoto++ Hinode CaIIH + Hida Halpha
% RR but the moustache extent would similarly be Thomson and H&K
% thermal widths themselves are narrower than Halpha (atomic mass).
% The fibrils obscure the core just as for Halpha (but much darker and
% less contrast because H&K have no J buildup below).  Note that the
% moustache resonance scattering in H&K is coherent.  So an electron
% can bounce it out of the core like a halo in the spectrum, resonance
% scattering gives a spatial halo.

Comment regarding EB fallacies: Severny
\citeads{1956Obs....76..241S} % Severny
\citeads{1957SvA.....1..668S}, % Severny review moustaches
recognizing that the moustache extent cannot be explained by Stark
broadening, invoked an EB scenario of nuclear explosion and
relativistic particle beams.
This way of thinking became a school of thought in which flares and
moustaches are regarded as similar phenomena, often mentioned
together, in which the photosphere is hit from above by particle beams
due to reconnection in the upper atmosphere.
Such beams would then explain the lack of \Halpha\ line-core
brightening by passing through and not affecting the \Halpha\
chromosphere.
Linear line-core polarization attributed to the beam impact became the
diagnostic for this notion
\citeads{1985SoPh...98..159B} % Babin+Koval polarization flares+moustaches
\citeads{1986SoPh..103...11F} % Firstova lin pol moustaches
\citeads{1998A&A...332..761D} % Ding+Henoux+Fang moustache impact beam
\citeads{1998A&A...337..294H} % Henoux etal proton beam
\citeads{1998RuPhJ..41.1258K} % Kazantsev+Firstova+Kashapova++
\citeads{2001ApJ...550L.105K} % Kosovichev+Zharkova flare with impact EBs
\citeads{2002ARep...46..918K} % Kashapova EB impact polarization
\citeads{2005A&A...431.1075Z} % Zharkova+Kashapova
\citeads{2006ApJ...643.1325F} % Fang++ Themis data + electron-beam models
\citeads{2011BCrAO.107...36B}. % Babin+Koval polarization flares, moustaches
%% (and more from the Crimea not served via ADS).  
In our opinion these authors failed to recognize that EBs are
photospheric and occur without anything happening overhead.
We note that in the example of
\citeads{1985SoPh...98..159B} % Babin+Koval polarization flares+moustaches
the polarization is not cospatial with the moustache, and that the
excess emission profiles of
\citeads{1986SoPh..103...11F} % Firstova lin pol moustaches
have no dip at line center.
We suspect that the measured line core polarizations stem from flaring
arch filaments, possibly through the mechanism of
\citeads{1983SoPh...86..115H}. % Henoux++ linear polarisation flares

% RR ?? hoe verklaren we linear Halpha core polarisatie?
% {2005A&A...431.1075Z} Zharkova+Kashapova: "polarisation only in
% cores" maar in cores zie je fibrils.  Eenvoudig linear polarisation
% van fibril transverse component?  Maar die moet je dan overal zien,
% niet alleen op EB locaties.  
% Firstova's emission profiles hebben geen line-center dip dus line-center
% emission dus waren flaring arches?  Polarisatie daarvan valt
% goed te verklaren met Henoux++

%%%%%%%%%%%%%%%%%%%%%%%%%%%%%%%%%%%%%%%%%%%%%%%%%%%%%%%%%%%%%%%%%%%%%%%%%%%%
\section{Ellerman bomb occurrence}
%%%%%%%%%%%%%%%%%%%%%%%%%%%%%%%%%%%%%%%%%%%%%%%%%%%%%%%%%%%%%%%%%%%%%%%%%%%%
The SST samples in Fig.~\ref{fig:EB2} include magnetograms that show
examples of what we find to be characteristic behavior in these data.
Small white
%## reversed contrast now
opposite-polarity Stokes-V patches produce EB flames in the \Halpha\
wing when colliding at relatively high speed with larger black
patches, and then vanish.
This is good evidence, similar to
\citeads{2008PASJ...60..577M} % Matsumoto++ Hinode CaIIH + Hida Halpha
but at higher resolution and with better statistics, that EBs mark
strong-field cancelation and supports the notion that the bright EB
flames are caused by photospheric reconnection
\citeads{1995Natur.375...42Y} % Yokoyama and Shibata reconnection
\citeads{1999ApJ...515..435L} % Litvinenko reconnection at photosphere
\citeads{2002ApJ...575..506G} % Georgoulis++ Flare Genesis
\citeads{2007ApJ...657L..53I} % Isobe+Tripathi+Archontis
\citeads{2008PASJ...60...95M} % Matsumoto++
\citeads{2009A&A...508.1469A} % Archontis+Hood
\citeads{2010PASJ...62..879H} % Hashimoto++ substructure EBs
\citeads{2011ApJ...736...71W}. % Watanabe++ EB1
The gasdynamical instability proposed by
\citeads{1996SoPh..168..105D} % Diver++ (John Brown) funny EB mechanism
seems a less likely explanation
%##
since it does not have field emergence as necessary condition.

Strong opposite-polarity field emergence may happen with small-scale
$\cap$ loop shape in classical $\Omega$ emergence
\citeads{1985SoPh..100..397Z} % Zwaan AR emergence
or with additional $\cup$ loop shape in serpentine emergence from the
Parker instability \citeads{1966ApJ...145..811P} % Parker instability
 \citeads{1992ApJS...78..267N}, % Nozawa++ undulatory emergence 
with connecting arch filaments
\citeads{1999ApJ...527..435S} % Strous+Zwaan
accompanied by dipped fields and bald patches
\citeads{2002SoPh..209..119B} % Bernasconi++ Flare Genesis
\citeads{2004ApJ...614.1099P} % Pariat++
\citeads{2006AdSpR..38..902P} % Pariat++ Flare Genesis, Omega emergence
\citeads{2008ApJ...684..736W} % Watanabe++ Hida spectra EFR
\citeads{2008ApJ...687.1373C} % Cheung++ undulatory emergence 
\citeads{2009ApJ...701.1911P} % Pariat++ serpentine emergence
\citeads{2009A&A...508.1469A} % Archontis+Hood
%%\citeads{2011PASJ...63.1047O} % Otsuji++ %RR geen goede ref, out
\citeads{2012ASPC..455..177P} % Pariat++ serpetine emergence
(see cartoons in
\citeads{2008ApJ...684..736W} % Watanabe++ Hida spectra EFR
%% \citeads{2010mcia.conf..505S} % Schmieder+Pariat Evershed meeting
\citeads{2012ASPC..455..177P} % Pariat++ serpetine emergence
\citeads{2012EAS....55..115P}). % Pariat++
In this picture chromospheric convergence of $\cup$ loop sides is
thought to give reconnection in the line-tied regime
\citeads{2007ApJ...657L..53I} % Isobe+Tripathi+Archontis
\citeads{2012EAS....55..115P}, % Pariat++
but the EBs in our SST data occur at colliding photospheric flows in
the frozen-in regime (Fig.~\ref{fig:EB2}).

Comment regarding EB fads: serpentine patterning into
opposite-polarity pairs of the MCs in the moat flow around decaying
spots was already suggested by
\citeads{1973SoPh...28...61H}, % Harvey+Harvey
but since EBs are emerging-flux phenomena it is tempting to assume
serpentine emergence and $\cup$ loop patterning.
This scenario may well be valid, but convincing proof requires
long-duration wide-field imaging spectroscopy with spectral, spatial,
and temporal resolution at least comparable to our SST sequences.
Obviously an IRIS topic.
Photospheric feature tracking following
\citeads{1996A&A...306..947S} % Strous++
would be a good technique to analyze such data.
Likewise, the notion of bi-directional anemone-jet flows with
reconnection a few hundred km up is attractive but needs further
verification.

%%%%%%%%%%%%%%%%%%%%%%%%%%%%%%%%%%%%%%%%%%%%%%%%%%%%%%%%%%%%%%%%%%%%%%%%%%%%
\section{Conclusion: Ellerman bomb usage}
%%%%%%%%%%%%%%%%%%%%%%%%%%%%%%%%%%%%%%%%%%%%%%%%%%%%%%%%%%%%%%%%%%%%%%%%%%%%
EBs are the most spectacular solar photosphere phenomenon.
They seem especially informative as space-time markers of strong-field
reconnection events in the photosphere, better than searching for
bipolar cancelation in magnetograms.
Comparison of the SST and \href{http://hmi.stanford.edu}{HMI}
magnetograms in Fig.~\ref{fig:EB2} illustrates the need for superhigh
resolution and sensitivity in the latter approach and shows that even
at the high SST quality one would not be able to identify reconnection
events from magnetogram sequences alone.
Larger Stokes sensitivity is desirable but longer integration would
degrade the temporal and spatial resolution fatally.
Thus, EB detection seems a better way to locate small-scale
reconnection events in emerging flux regions.
Since the larger and brighter EBs are also visible in 1700\,\AA\
images even at the AIA resolution, the AIA database permits
monitoring active-region field re-configuration through EB
identification in large data volumes
\cite{Vissers++2013b}. % EB3
EBs may so become useful photospheric telltales in studying
chromospheric active-region field topology evolution and energy
loading.

\ack We thank R.~Rezaei for informative discussions and the referee
for valuable suggestions.
RJR acknowledges that the late C.~Zwaan often suggested EBs as
research topic, and thanks the Leids Kerkhoven-Bosscha Fonds for
travel support.
We made much use of NASA's Astrophysics Data System Bibliographic
Services.
The macro to make the citations above link to ADS (at least in the
arXiv preprint) was contributed by
\href{http://ftp.edpsciences.org/pub/aa/readme.txt}{EDP Sciences}.

%%%%%%%%%%%%%%%%%%%%%%%%%%%%%%%%%%%%%%%%%%%%%%%%%%%%%%%%%%%%%%%%%%%%%%%%%%%%
%%\spacing{0.95}  %RR ?? adapt to fill last page
%RR IoP edits from pdf file, fully their business
% \bibliographystyle{iopart-num}
% \bibliography{journals,rjrfiles,adsfiles} %RR rjr first for ADS repairs

\begin{thebibliography}{100}
\expandafter\ifx\csname url\endcsname\relax
  \def\url#1{{\tt #1}}\fi
\expandafter\ifx\csname urlprefix\endcsname\relax\def\urlprefix{URL }\fi
\providecommand{\eprint}[2][]{\url{#2}}

\parskip=-0.5ex

\bibitem{1917ApJ....46..298E}
{Ellerman} F 1917 {\em \apj\/} {\bf 46} 298

\bibitem{1974soch.book.....B}
{Bray} R~J and {Loughhead} R~E 1974 {\em {The solar chromosphere}\/}
Chapman \& Hall, London

\bibitem{2002SoPh..209..119B}
{Bernasconi} P~N, {Rust} D~M, {Georgoulis} M~K and {Labonte} B~J 2002 {\em
  \solphys\/} {\bf 209} 119

\bibitem{2002ApJ...575..506G}
{Georgoulis} M~K, {Rust} D~M, {Bernasconi} P~N and {Schmieder} B 2002 {\em
  \apj\/} {\bf 575} 506

\bibitem{2002ESASP.506..911S}
{Schmieder} B, {Pariat} E, {Aulanier} G, {Georgoulis} M~K, {Rust} D~M and
  {Bernasconi} P~N 2002 in {\em Solar Variability: From Core to Outer Frontiers\/}
  ({\em ESA Special Pub.\/} 506) ed {Wilson} A 911

\bibitem{2004ApJ...601..530S}
{Schmieder} B, {Rust} D~M, {Georgoulis} M~K, {D{\'e}moulin} P and {Bernasconi}
  P~N 2004 {\em \apj\/} {\bf 601} 530

\bibitem{2004ApJ...614.1099P}
{Pariat} E, {Aulanier} G, {Schmieder} B, {Georgoulis} M~K, {Rust} D~M and
  {Bernasconi} P~N 2004 {\em \apj\/} {\bf 614} 1099

\bibitem{2006AdSpR..38..902P}
{Pariat} E, {Aulanier} G, {Schmieder} B, {Georgoulis} M~K, {Rust} D~M and
  {Bernasconi} P~N 2006 {\em Adv.\ Space Research\/} {\bf 38} 902

\bibitem{2011ApJ...736...71W}
{Watanabe} H, {Vissers} G, {Kitai} R, {Rouppe van der Voort} L and {Rutten} R~J
  2011 {\em \apj\/} {\bf 736} 71

\bibitem{Vissers++2013a}
Vissers G, {Rouppe van der Voort} L~H~M and Rutten R~J 2013 {\em ApJ\/}
  submitted

\bibitem{2009A&A...499..301V}
{Vitas} N, {Viticchi{\`e}} B, {Rutten} R~J and {V{\"o}gler} A 2009 {\em \aap\/}
  {\bf 499} 301

\bibitem{1969SoPh....9..347S}
{Sheeley} Jr N~R 1969 {\em \solphys\/} {\bf 9} 347

\bibitem{1971IAUS...43..329V}
{Vrabec} D 1971 in {\em Solar Magnetic Fields\/} ({\em IAU Symposium\/} vol~43) ed
  {Howard} R 329

\bibitem{1973SoPh...28...61H}
{Harvey} K and {Harvey} J 1973 {\em \solphys\/} {\bf 28} 61

\bibitem{1974IAUS...56..201V}
{Vrabec} D 1974 in {\em Chromospheric Fine Structure\/} ({\em IAU Symposium\/}
  vol~56) ed {Athay} R~G 201

\bibitem{1987SoPh..112..295M}
{Muller} R and {Mena} B 1987 {\em \solphys\/} {\bf 112} 295

\bibitem{2005ApJ...635..659H}
{Hagenaar} H~J and {Shine} R~A 2005 {\em \apj\/} {\bf 635} 659

\bibitem{1976SoPh...50..269S}
{Spruit} H~C 1976 {\em \solphys\/} {\bf 50} 269

\bibitem{1977PhDT.......237S}
{Spruit} H~C 1977 {\em {Magnetic flux tubes and transport of heat in the
  convection zone of the Sun.}\/} Ph.D. Thesis Utrecht University

\bibitem{1981NASSP.450..385S} {Spruit} H~C 1981 in {\em The Sun as a
    Star\/} ed Jordan S {\em NASA Special Pub.\/} {\bf 450} 385

\bibitem{1984A&A...139..426D}
{Deinzer} W, {Hensler} G, {Sch{\"{u}}ssler} M and {Weisshaar} E 1984 {\em
  \aap\/} {\bf 139} 426

\bibitem{1988A&A...194..257K}
{Kn{\"{o}}lker} M, {Sch{\"{u}}ssler} M and {Weisshaar} E 1988 {\em \aap\/} {\bf
  194} 257

\bibitem{1991sia..book..890S} {Spruit} H~C, {Sch{\"{u}}ssler} M and
  {Solanki} S~K 1991 in {\em Solar Interior and Atmosphere\/} Univ
  Arizona Press 890

\bibitem{1992A&A...262L..29S}
{Solanki} S~K and {Brigljevic} V 1992 {\em \aap\/} {\bf 262} L29

\bibitem{1993A&A...268..736B}
{B{\"{u}}nte} M, {Solanki} S~K and {Steiner} O 1993 {\em \aap\/} {\bf 268}
  736

\bibitem{1993SSRv...63....1S}
{Solanki} S~K 1993 {\em \ssr\/} {\bf 63} 1

\bibitem{2009A&A...504..595Y}
{Yelles Chaouche} L, {Solanki} S~K and {Sch{\"u}ssler} M 2009 {\em \aap\/} {\bf
  504} 595

\bibitem{2004A&A...428..613B}
{Berger} T~E, {Rouppe van der Voort} L~H~M, {L{\"o}fdahl} M~G, {Carlsson} M,
  {Fossum} A, {Hansteen} V~H, {Marthinussen} E, {Title} A and {Scharmer} G 2004
  {\em \aap\/} {\bf 428} 613

\bibitem{2005A&A...435..327R}
{Rouppe van der Voort} L~H~M, {Hansteen} V~H, {Carlsson} M, {Fossum} A,
  {Marthinussen} E, {van Noort} M~J and {Berger} T~E 2005 {\em \aap\/} {\bf
  435} 327

\bibitem{1996ApJ...463..797T}
{Title} A~M and {Berger} T~E 1996 {\em \apj\/} {\bf 463} 797

\bibitem{1987ApJ...314..808L}
{Livingston} W and {Wallace} L 1987 {\em \apj\/} {\bf 314} 808

\bibitem{2007ApJ...657.1137L}
{Livingston} W, {Wallace} L, {White} O~R and {Giampapa} M~S 2007 {\em \apj\/}
  {\bf 657} 1137

\bibitem{2006A&A...449.1209L}
{Leenaarts} J, {Rutten} R~J, {S{\"u}tterlin} P, {Carlsson} M and {Uitenbroek} H
  2006 {\em \aap\/} {\bf 449} 1209

\bibitem{2006A&A...452L..15L}
{Leenaarts} J, {Rutten} R~J, {Carlsson} M and {Uitenbroek} H 2006 {\em \aap\/}
  {\bf 452} L15

\bibitem{2012RSPTA.370.3129R}
{Rutten} R~J 2012 {\em Roy.\ Soc.\ London Phil.\ Trans.\
  Series A\/} {\bf 370} 3129

\bibitem{1972SoPh...22..344S}
{Schoolman} S~A 1972 {\em \solphys\/} {\bf 22} 344

\bibitem{2012A&A...540A..86R}
{Rutten} R~J and {Uitenbroek} H 2012 {\em \aap\/} {\bf 540} A86

\bibitem{2012ApJ...749..136L}
{Leenaarts} J, {Carlsson} M and {Rouppe van der Voort} L 2012 {\em \apj\/} {\bf
  749} 136

\bibitem{1972ApJ...178L..85Z}
{Zirin} H and {Stein} A 1972 {\em \apjl\/} {\bf 178} L85

\bibitem{2001A&A...375..617C}
{Christopoulou} E~B, {Georgakilas} A~A and {Koutchmy} S 2001 {\em \aap\/} {\bf
  375} 617

\bibitem{2002ApJ...572..626C}
{Carlsson} M and {Stein} R~F 2002 {\em \apj\/} {\bf 572} 626

\bibitem{2007A&A...473..625L}
{Leenaarts} J, {Carlsson} M, {Hansteen} V and {Rutten} R~J 2007 {\em \aap\/}
  {\bf 473} 625

\bibitem{2007A&A...471.1035Z}
{Zhang} J, {Solanki} S~K, {Woch} J and {Wang} J 2007 {\em \aap\/} {\bf 471}
1035

\bibitem{2011ApJ...730L..24K}
{Kato} Y, {Steiner} O, {Steffen} M and {Suematsu} Y 2011 {\em \apjl\/} {\bf
  730} L24

\bibitem{2011A&A...531A..17R}
{Rutten} R~J, {Leenaarts} J, {Rouppe van der Voort} L~H~M, {De Wijn} A~G,
  {Carlsson} M and {Hansteen} V 2011 {\em \aap\/} {\bf 531} A17

\bibitem{2008ApJ...687L.131B}
{Bonet} J~A, {M{\'a}rquez} I, {S{\'a}nchez Almeida} J, {Cabello} I and
  {Domingo} V 2008 {\em \apjl\/} {\bf 687} L131

\bibitem{2007A&A...476..971J}
{Jess} D~B, {McAteer} R~T~J, {Mathioudakis} M, {Keenan} F~P, {Andic} A and
  {Bloomfield} D~S 2007 {\em \aap\/} {\bf 476} 971

\bibitem{2005A&A...441.1183D}
{De Wijn} A~G, {Rutten} R~J, {Haverkamp} E~M~W~P and {S{\"u}tterlin} P 2005
  {\em \aap\/} {\bf 441} 1183

\bibitem{1956Obs....76..241S}
{Severny} A~B 1956 {\em The Observatory\/} {\bf 76} 241

\bibitem{1957SvA.....1..668S}
{Severnyi} A~B 1957 {\em \sovast\/} {\bf 1} 668

\bibitem{1968IAUS...35..293B}
{Bruzek} A 1968 in {\em Structure and Development of Solar Active Regions\/} ({\em
  IAU Symposium\/} vol~35) ed {Kiepenheuer} K~O 293

\bibitem{1970SoPh...14..112T}
{Turon} P~J and {L{\'e}na} P~J 1970 {\em \solphys\/} {\bf 14} 112

\bibitem{1972SoPh...26...94B}
{Bruzek} A 1972 {\em \solphys\/} {\bf 26} 94

\bibitem{1987SoPh..108..227Z}
{Zachariadis} T~G, {Alissandrakis} C~E and {Banos} G 1987 {\em \solphys\/} {\bf
  108} 227

\bibitem{1991A&A...251..675S}
{Stellmacher} G and {Wiehr} E 1991 {\em \aap\/} {\bf 251} 675

\bibitem{1995A&A...296..567D}
{Denker} C, {de Boer} C~R, {Volkmer} R and {Kneer} F 1995 {\em \aap\/} {\bf
  296} 567

\bibitem{1997A&A...323..599D}
{Denker} C 1997 {\em \aap\/} {\bf 323} 599

\bibitem{1998SoPh..182..381N}
{Nindos} A and {Zirin} H 1998 {\em \solphys\/} {\bf 182} 381

\bibitem{1999ESASP.448..279G}
{Georgakilas} A~A, {Christopoulou} E~B and {Koutchmy} S 1999 in {\em Magnetic
  Fields and Solar Processes\/} ({\em ESA Special Pub.\/} 448) ed
  {Wilson} A and {et al} 279

\bibitem{2013SoPh..283..307N}
{Nelson} C~J, {Doyle} J~G, {Erd{\'e}lyi} R, {Huang} Z, {Madjarska} M~S,
  {Mathioudakis} M, {Mumford} S~J and {Reardon} K 2013 {\em \solphys\/} {\bf
  283} 307

\bibitem{1982SoPh...79...77K}
{Kurokawa} H, {Kawaguchi} I, {Funakoshi} Y and {Nakai} Y 1982 {\em \solphys\/}
  {\bf 79} 77

\bibitem{2007A&A...473..279P}
{Pariat} E, {Schmieder} B, {Berlicki} A, {Deng} Y, {Mein} N, {L{\'o}pez Ariste}
  A and {Wang} S 2007 {\em \aap\/} {\bf 473} 279

\bibitem{2010PASJ...62..879H}
{Hashimoto} Y, {Kitai} R, {Ichimoto} K, {Ueno} S, {Nagata} S, {Ishii} T~T,
  {Hagino} M, {Komori} H, {Nishida} K, {Matsumoto} T, {Otsuji} K, {Nakamura} T,
  {Kawate} T, {Watanabe} H and {Shibata} K 2010 {\em \pasj\/} {\bf 62} 879

\bibitem{1963Obs....83...37H}
{Harvey} J~W 1963 {\em The Observatory\/} {\bf 83} 37

\bibitem{2008ApJ...684..736W}
{Watanabe} H, {Kitai} R, {Okamoto} K, {Nishida} K, {Kiyohara} J, {Ueno} S,
  {Hagino} M, {Ishii} T~T and {Shibata} K 2008 {\em \apj\/} {\bf 684} 736

\bibitem{1971SoPh...18..347G}
{Gingerich} O, {Noyes} R~W, {Kalkofen} W and {Cuny} Y 1971 {\em \solphys\/}
  {\bf 18} 347

\bibitem{1981ApJS...45..635V}
{Vernazza} J~E, {Avrett} E~H and {Loeser} R 1981 {\em \apjs\/} {\bf 45}
  635

\bibitem{1993ApJ...406..319F}
{Fontenla} J~M, {Avrett} E~H and {Loeser} R 1993 {\em \apj\/} {\bf 406}
  319

\bibitem{2008ApJS..175..229A}
{Avrett} E~H and {Loeser} R 2008 {\em \apjs\/} {\bf 175} 229

\bibitem{2010ApJ...709.1362L}
{Leenaarts} J, {Rutten} R~J, {Reardon} K, {Carlsson} M and {Hansteen} V 2010
  {\em \apj\/} {\bf 709} 1362

\bibitem{1970SoPh...11..276K}
{Koval} A~N and {Severny} A~B 1970 {\em \solphys\/} {\bf 11} 276

\bibitem{1962AuJPh..15..327B}
{Beckers} J~M 1962 {\em Austr.\ J.\ Physics\/} {\bf 15} 327

\bibitem{1990A&A...233..207D}
{Dere} K~P, {Schmieder} B and {Alissandrakis} C~E 1990 {\em \aap\/} {\bf 233}
  207

\bibitem{2008A&A...491L...5T}
{Teriaca} L, {Curdt} W and {Solanki} S~K 2008 {\em \aap\/} {\bf 491} L5

\bibitem{1997A&A...322..653D}
{Dara} H~C, {Alissandrakis} C~E, {Zachariadis} T~G and {Georgakilas} A~A 1997
  {\em \aap\/} {\bf 322} 653

\bibitem{1973SoPh...28...95R}
{Roy} J~R 1973 {\em \solphys\/} {\bf 28} 95

\bibitem{1973SoPh...30..449R}
{Roy} J~R and {Leparskas} H 1973 {\em \solphys\/} {\bf 30} 449

\bibitem{1979SoPh...63..353C}
{Carlqvist} P 1979 {\em \solphys\/} {\bf 63} 353

\bibitem{1982SoPh...77..121S}
{Shibata} K, {Nishikawa} T, {Kitai} R and {Suematsu} Y 1982 {\em \solphys\/}
  {\bf 77} 121

\bibitem{2007ApJ...657L..53I}
{Isobe} H, {Tripathi} D and {Archontis} V 2007 {\em \apjl\/} {\bf 657} L53

\bibitem{2008PASJ...60...95M}
{Matsumoto} T, {Kitai} R, {Shibata} K, {Otsuji} K, {Naruse} T, {Shiota} D and
  {Takasaki} H 2008 {\em \pasj\/} {\bf 60} 95

\bibitem{2007ASPC..369..113S}
{Shimizu} T, {Mart{\'{\i}}nez-Pillet} V, {Collados} M, {Ruiz-Cobo} B, {Centeno}
  R, {Beck} C and {Katsukawa} Y 2007 in {\em New Solar Physics with Solar-B
  Mission\/} ({\em \aspcs\/} vol
  369) ed {Shibata} K, {Nagata} S and {Sakurai} T 113

\bibitem{2007Sci...318.1591S}
{Shibata} K, {Nakamura} T, {Matsumoto} T, {Otsuji} K, {Okamoto} T~J,
  {Nishizuka} N, {Kawate} T, {Watanabe} H, {Nagata} S, {UeNo} S, {Kitai} R,
  {Nozawa} S, {Tsuneta} S, {Suematsu} Y, {Ichimoto} K, {Shimizu} T, {Katsukawa}
  Y, {Tarbell} T~D, {Berger} T~E, {Lites} B~W, {Shine} R~A and {Title} A~M 2007
  {\em Science\/} {\bf 318} 1591

\bibitem{2010PASJ...62..901M}
{Morita} S, {Shibata} K, {Ueno} S, {Ichimoto} K, {Kitai} R and {Otsuji} K~I
  2010 {\em \pasj\/} {\bf 62} 901

\bibitem{2011ApJ...731...43N}
{Nishizuka} N, {Nakamura} T, {Kawate} T, {Singh} K~A~P and {Shibata} K 2011
  {\em \apj\/} {\bf 731} 43

\bibitem{2011CEAB...35..181H}
{Herlender} M and {Berlicki} A 2011 {\em Central European Astrophys.\
  Bull.\/} {\bf 35} 181

\bibitem{2000ApJ...544L.157Q}
{Qiu} J, {Ding} M~D, {Wang} H, {Denker} C and {Goode} P~R 2000 {\em \apjl\/}
  {\bf 544} L157

\bibitem{2006ApJ...643.1325F}
{Fang} C, {Tang} Y~H, {Xu} Z, {Ding} M~D and {Chen} P~F 2006 {\em \apj\/} {\bf
  643} 1325

\bibitem{2006SoPh..235...75S}
{Socas-Navarro} H, {Mart{\'{\i}}nez Pillet} V, {Elmore} D, {Pietarila} A,
  {Lites} B~W and {Manso Sainz} R 2006 {\em \solphys\/} {\bf 235} 75

\bibitem{2008PASJ...60..577M}
{Matsumoto} T, {Kitai} R, {Shibata} K, {Nagata} S, {Otsuji} K, {Nakamura} T,
  {Watanabe} H, {Tsuneta} S, {Suematsu} Y, {Ichimoto} K, {Shimizu} T,
  {Katsukawa} Y, {Tarbell} T~D, {Lites} B~W, {Shine} R~A and {Title} A~M 2008
  {\em \pasj\/} {\bf 60} 577

\bibitem{2010MmSAI..81..646B}
{Berlicki} A, {Heinzel} P and {Avrett} E~H 2010 {\em \memsai\/} {\bf 81} 646

\bibitem{2005ApJ...625..556F}
{Fossum} A and {Carlsson} M 2005 {\em \apj\/} {\bf 625} 556

\bibitem{2002ApJ...574.1074S}
{Shimizu} T, {Shine} R~A, {Title} A~M, {Tarbell} T~D and {Frank} Z 2002 {\em
  \apj\/} {\bf 574} 1074

\bibitem{2010ApJ...724.1083G}
{Guglielmino} S~L, {Bellot Rubio} L~R, {Zuccarello} F, {Aulanier} G, {Vargas
  Dom{\'{\i}}nguez} S and {Kamio} S 2010 {\em \apj\/} {\bf 724} 1083

\bibitem{2009ApJ...701..253M}
{Madjarska} M~S, {Doyle} J~G and {De Pontieu} B 2009 {\em \apj\/} {\bf 701}
  253

\bibitem{2010CEAB...34...65H}
{Herlender} M and {Berlicki} A 2010 {\em Central European Astrophys.\
  Bull.\/} {\bf 34} 65

\bibitem{2010ApJ...712L.111J}
{Jess} D~B, {Mathioudakis} M, {Browning} P~K, {Crockett} P~J and {Keenan} F~P
  2010 {\em \apjl\/} {\bf 712} L111

\bibitem{1996SoPh..165..155K}
{Kitai} R and {Muller} R 1996 {\em \solphys\/} {\bf 165} 155

\bibitem{Vissers++2013b}
Vissers G, {Rouppe van der Voort} L~H~M and Rutten R~J 2013   in preparation

\bibitem{2003rtsa.book.....R}
{Rutten} R~J 2003 {\em {Radiative Transfer in Stellar Atmospheres}\/}
Lecture Notes Utrecht University

\bibitem{1983SoPh...87..135K}
{Kitai} R 1983 {\em \solphys\/} {\bf 87} 135

\bibitem{1968mmsf.conf..109E}
{Engvold} O and {Maltby} P 1968 in {\em Mass Motions in Solar Flares and Related
  Phenomena\/}  (Procs 9th Nobel Symposium) ed {Oehman} Y 109

\bibitem{1985SoPh...98..159B}
{Babin} A~N and {Koval} A~N 1985 {\em \solphys\/} {\bf 98} 159

\bibitem{1986SoPh..103...11F}
{Firstova} N~M 1986 {\em \solphys\/} {\bf 103} 11

\bibitem{1998A&A...332..761D}
{Ding} M~D, {Henoux} J~C and {Fang} C 1998 {\em \aap\/} {\bf 332} 761

\bibitem{1998A&A...337..294H}
{Henoux} J~C, {Fang} C and {Ding} M~D 1998 {\em \aap\/} {\bf 337} 294

\bibitem{1998RuPhJ..41.1258K}
{Kazantsev} S~A, {Firstova} N~M, {Kashapova} L~K, {Bulatov} A~V and
  {Petrashen'} A~G 1998 {\em Russian Phys.\ J.\/} {\bf 41} 1258

\bibitem{2001ApJ...550L.105K}
{Kosovichev} A~G and {Zharkova} V~V 2001 {\em \apjl\/} {\bf 550} L105

\bibitem{2002ARep...46..918K}
{Kashapova} L~K 2002 {\em Astron.\ Reports\/} {\bf 46} 918

\bibitem{2005A&A...431.1075Z}
{Zharkova} V~V and {Kashapova} L~K 2005 {\em \aap\/} {\bf 431} 1075

\bibitem{2011BCrAO.107...36B}
{Babin} A~N and {Koval} A~N 2011 {\em Bulletin Crimean Astrophys.\
  Obs.\/} {\bf 107} 36

\bibitem{1983SoPh...86..115H}
{Henoux} J~C, {Chambe} G, {Heristchi} D, {Semel} M, {Woodgate} B, {Shine} D and
  {Beckers} J 1983 {\em \solphys\/} {\bf 86} 115

\bibitem{1995Natur.375...42Y}
{Yokoyama} T and {Shibata} K 1995 {\em \nat\/} {\bf 375} 42

\bibitem{1999ApJ...515..435L}
{Litvinenko} Y~E 1999 {\em \apj\/} {\bf 515} 435

\bibitem{2009A&A...508.1469A}
{Archontis} V and {Hood} A~W 2009 {\em \aap\/} {\bf 508} 1469

\bibitem{1996SoPh..168..105D}
{Diver} D~A, {Brown} J~C and {Rust} D~M 1996 {\em \solphys\/} {\bf 168}
  105

\bibitem{1985SoPh..100..397Z}
{Zwaan} C 1985 {\em \solphys\/} {\bf 100} 397

\bibitem{1966ApJ...145..811P}
{Parker} E~N 1966 {\em \apj\/} {\bf 145} 811

\bibitem{1992ApJS...78..267N}
{Nozawa} S, {Shibata} K, {Matsumoto} R, {Sterling} A~C, {Tajima} T, {Uchida} Y,
  {Ferrari} A and {Rosner} R 1992 {\em \apjs\/} {\bf 78} 267

\bibitem{1999ApJ...527..435S}
{Strous} L~H and {Zwaan} C 1999 {\em \apj\/} {\bf 527} 435

\bibitem{2008ApJ...687.1373C}
{Cheung} M~C~M, {Sch{\"u}ssler} M, {Tarbell} T~D and {Title} A~M 2008 {\em
  \apj\/} {\bf 687} 1373

\bibitem{2009ApJ...701.1911P}
{Pariat} E, {Masson} S and {Aulanier} G 2009 {\em \apj\/} {\bf 701} 1911

\bibitem{2012ASPC..455..177P} {Pariat} E, {Masson} S and {Aulanier} G
  2012 in {\em 4th Hinode Science Meeting: Unsolved Problems and Recent
    Insights\/} ({\em \aspcs\/} 455) ed {Bellot Rubio} L, {Reale}
  F and {Carlsson} M 177

\bibitem{2012EAS....55..115P}
{Pariat} E, {Schmieder} B, {Masson} S and {Aulanier} G 2012 in {\em EAS
  Publications Series\/} 55) ed
  {Faurobert} M, {Fang} C and {Corbard} T 115

\bibitem{1996A&A...306..947S}
{Strous} L~H, {Scharmer} G, {Tarbell} T~D, {Title} A~M and {Zwaan} C 1996 {\em
  \aap\/} {\bf 306} 947

\end{thebibliography}
%% \input{eb-pc-clean.bbl}

%% file: cleaned uo version 2nd iteration
%% last: Apr  5 2013  Rob Rutten  LA Deil

\providecommand{\newblock}{}

\end{document}